\documentstyle[11pt,newpasp,twoside,epsf]{article}
\begin{document}
\title{T Tauri Multiple Systems}
 \author{Alexis Brandeker}
\affil{Stockholm~Observatory, AlbaNova~University~Centre,
SE-106~91~Stockholm, Sweden}

\begin{abstract}
New high-resolution adaptive optics systems provide an unprecedentedly
detailed view of nearby star forming regions. In particular, young
nearby T~Tauri stars can be probed at much smaller physical scales
(a few AU) than possible just a decade ago (several tens of AU). Of
major importance is closing the sensitivity gap between imaging and
spectral surveys for stellar companions. This allows for
1)~calibration of pre-main-sequence evolutionary tracks by obtaining
accurate dynamical masses, 
2)~resolving confusion problems arising by placing unresolved systems in
colour-magnitude diagrams, and 
3)~well defined and determined multiplicity fractions of young stellar
systems, important for discriminating star formation scenarios. This article
briefly reviews the current status of high resolution imaging of T~Tauri
multiple systems, and what we can expect to learn from them
in the near future.
\end{abstract}

\section{Introduction}
Most stars, especially pre-main-sequence (PMS) stars, are part of multiple
systems (Duch\^{e}ne 1999; reviews by Tom Greene and Ralf Launhardt, this volume).
Since single stars are only in minority, we need to
understand how multiple systems are formed in order to understand star formation in
general. Because it is currently not possible to deduce a star formation theory from
first principles without introducing unrealistic assumptions, observations of young
systems are essential for differentiating between possible hypothetical scenarios.

Over the last years, some hundred nearby T~Tauri stars have been imaged with adaptive
optics, and before that some hundred were studied by speckle interferometry and lunar
occultations. The three examples in Fig.~1 show adaptive optics (AO) imaging of three
different multiple T~Tauri stars put to the same spatial scale. They may look surprisingly
similar, but are actually typical in two ways: a large fraction of the systems we image
are not only binary, but often triple or quadruple, and multiple stars are mostly
hierarchically arranged.

Two major questions that can be addressed by high angular resolution observations
of multiple stars are:
\begin{itemize}
\item[$\bullet$] How does the early stellar evolution depend on mass (and metallicity)?
\item[$\bullet$] Does the multiplicity frequency for stellar systems evolve in time?
\end{itemize}
In the following sections I will briefly describe methods of finding answers
to these questions, with some emphasis on the contributions from modern
AO systems. I will also review a few representative recent results
in the area, and discuss what we may expect in the near future.

\begin{figure}
\plotone{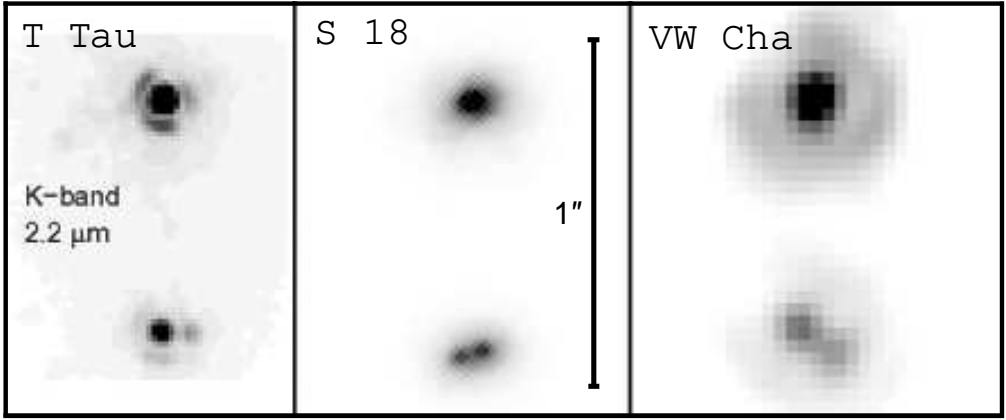}
\caption{Three different triple T~Tauri systems with a similar appearance: T~Tauri, 
S~18 (in the MBM~12 young star association) and VW~Chamaeleontis. The observations,
here put to the same angular scale, are from Duch\^{e}ne, Ghez, \& McCabe (2002),
Brandeker, Jayawardhana, \& Najita (2003), and Brandeker et al.\ (2001) respectively.
T~Tau was observed in $K$ with Keck AO, S~18 also with Keck AO but in $H$, and
VW~Cha in $K$ with ADONIS at the 3.6\,m telescope at ESO, La Silla.
\label{fig-1}}
\end{figure}

\section{Calibrating PMS evolutionary tracks}
Historically, binaries have played a major role in the successful theory
of main sequence (MS) stellar structure. The importance of binaries in this context
lies in the possibility of direct determination of stellar dynamical masses
by following the orbital motion. For PMS stars, a problem has been that the
nearest star forming regions are on the order of 100--150~pc distant. With
observations limited by a typical seeing disc of 1\arcsec, this implies
a projected physical resolution of 100--150~AU. Even for massive systems, the
orbital period at the resolution separation is expected to be on the order of
hundreds to thousands of years, making dynamical mass estimates impractical.
Tight short-period binaries may still be found spectroscopically, since 
spectroscopic searches are limited by flux and not spatial scale. Without
spatial information, however, only the relative masses of the stars in a
binary can be deduced, unless the inclination of the orbit is known by other
means.

\begin{figure}
\plotone{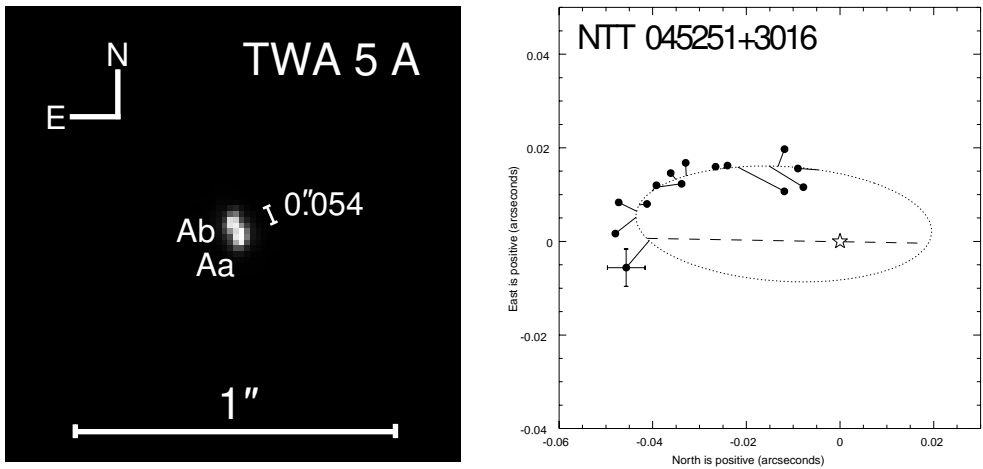}
\caption{ ({\it a, left panel}) The tight inner 3~AU binary TWA~5A of the TW~Hydrae
association. Expected period of binary is 5--10~yr. From Brandeker,
Jayawardhana \& Najita (2003). ({\it b, right panel}) Orbital solution to PMS binary
NTT~045251+3016, making use of both spatially resolved observations and spectroscopic
radial velocity measurements. From Steffen et al.\ (2001).
\label{fig-2}}
\end{figure}

\subsection{Dynamical mass determinations}
Clearly observations at higher angular resolutions are essential for
dynamical mass determinations. AO systems on modern 8--10~m facilities
regularly achieve diffraction limited imaging in the near infrared (NIR),
corresponding to angular resolutions of 30--50~mas. At the distance of
recently discovered nearby young clusters, like the TW~Hydrae
association (TWA) at a distance of $\sim$55~pc, this translates to a
projected physical scale of merely $\sim$2~AU. An example of a tight
binary close to the resolution limit of the Keck telescope AO is shown in
Fig.~2a, displaying TWA~5A with an estimated projected separation of 3~AU. By
tracking the orbital motion of such tight binaries, it is possible to estimate
a dynamical mass within just a few years. NTT~045251+3016 is a recent example
of a binary dynamical mass estimate by Steffen et al.\ 2001, with their orbital
solution shown in Fig.~2b. Using the Fine Guidance Sensor of the Hubble
Space Telescope, they were able to reach a relative positional accuracy of 4~mas.
Following the orbit during 3 years, a little less than half of the estimated orbital
period, and using additional spectroscopically measured radial velocity differences
between the two components, individual masses for the stars could be established as
$1.45\pm0.19\,\rm{M_{\sun}}$ and $0.81\pm0.09\,\rm{M_{\sun}}$. Possibly surprising, 
a dynamical {\it distance} $d=144.8\pm8.3$~pc was also derived by combining astrometric
and spectroscopic data, independent of any paralactic distance. Schaefer et al.\ (2003)
report three more systems with the system mass dynamically estimated,
and show that meaningful dynamical mass estimates can be obtained even though
the orbital elements remain very uncertain, as first noted by Eggen (1967). 

Other methods that have been employed to determine stellar dynamical masses involve
eclipsing binaries (e.g.\ Covino et al.\ 2000) and orbital motion of disk gas
(Simon, Dutrey, \& Guilloteau 2000). Both have their benefits and caveats:
Although eclipsing binaries provide very accurate masses, they require special
geometry (very small orbital inclination to the line of sight), and are consequently
comparatively rare. By studying disk gas motion also the mass of single stars can be
dynamically estimated. Gas motion is, however, more sensitive to non-gravitational
effects, such as the radiation pressure (see e.g. Olofsson, Liseau, \& Brandeker
2001).

Accurate dynamical mass estimates of young low-mass stars ($<1\,M_{\sun}$)
remain very rare, but given the recent advances in high angular resolution 
instrumentation, and a few years of orbital motion, the near future will no
doubt see the number of dynamically estimated masses multiply.

\subsection{Resolving confusion}
Another problem with seeing limited observations is that of confusion. Tight
multiple systems, like those in Fig.~1, are entirely contained in a seeing
disc, meaning that photometry will be measured on the system as a whole.
Placing the measurements in a Hertzsprung-Russell diagram, the whole system
will show an offset from the evolutionary tracks compared to the individual stars.
Faint infrared companions close to the primary may show up as an IR-excess in
the measured spectral energy distribution, difficult to distinguish from
e.g. a warm circumstellar disk. Spectroscopic searches for faint companions
are most sensitive to shorter periods, and are inefficient in finding binaries
with periods of a couple of years or more. This means physical separations of a
few AU or less, translating to some 40~mas or less at the typical distances of
100--150~pc. This is on the same order as the angular resolution of modern
AO systems, meaning that there no longer is a sensitivity gap where stellar
companions can hide. 

\section{Multiplicity frequency evolution}
Most MS stellar systems in the solar neighbourhood are multiple
(Duquennoy \& Mayor 1991), and for PMS systems the multiplicity fraction in general 
seems to be even higher (e.g.\ Reipurth \& Zinnecker 1993; Duch\^{e}ne 1999;
Barsony, Koresko, \& Matthews 2003).
How can this be, if we believe MS field stars to be evolved T~Tauri stars? I know 
of three proposed explanations:

\subsection{Selection effects}
Maybe the multiple fraction excess among PMS stars is due
to more sensitive searches (Ghez 1996). For instance, young low-mass companions are brighter
and more easily seen than their older counterparts, especially in the NIR where
most searches are conducted. However, a more detailed study by White \& Ghez (2001)
argues this to not be the case: we do see more companions among young stars,
even when looking at the same mass intervals. This point illuminates the importance
of finding and correcting observational biases. By closing the sensitivity gap
between spectral and spatial searches for companions, corrections for unseen
components can be put on a firmer ground.

\subsection{Regional differences}
Maybe different regions produce different multiplicity fractions, and the star
forming regions we observe just by chance happen to have higher multiplicity
fractions than the average. Indeed, different regions apparently do show different
multiplicity fractions (Duch\^{e}ne 1999). Possible parameters responsible for this
``environmental effect'' are metallicity, volume stellar densities and internal 
velocity dispersion (see also Woitas, Leinert, \& K\"{o}hler 2001; Zinnecker 2003).
The statistics are still quite poor, however, so the significance of variations 
among different regions remains low.

\subsection{Dynamical evolution}
Perhaps multiple systems evolve dynamically, and in the process lose companions,
either by stellar mergers or ejections (e.g.\ Reipurth 2000). One may also speculate
that orbital evolution make companions harder to detect. One way is by shrinking a
wide, easily resolved, binary to a tight, not so easily resolved binary, another
is to widen a moderately wide binary to a very wide (and loose) binary, which will
consequently make the companions hard to
identify as such (and also make the system more sensitive to disruption due to stellar
encounters). By increasing the angular resolution to the point where any stellar
companion tighter than the resolution limit would be picked up spectroscopically,
the hypothesis of unseen tight binaries can be ruled out. Very wide and loose
binaries may be more difficult to find, but can be addressed by multiple epoch
wide field studies of proper motions in combination with radial velocity
observations, to assess relative space motions.

 Detailed numerical studies of the early orbital evolution have produced
some quantitative predictions of the low-mass stellar population in young clusters
(Bates, Bonnell, \& Bromm 2002, 2003) that can be checked in the near future.
Companion ejection requires three-body interaction, and would thus mainly affect
the companion fraction defined as $cf = (2b + 3t + 4q + ...) / (s + b + t + q + ...)$,
 where $s$, $b$, $t$, $q$, etc.\ are the number of single stars, binaries, triples,
quadruples etc. The companion fraction is the average number of companions to any
primary. The multiplicity fraction $mf = (b + t + q + ...) / (s + b + t + q + ...)$,
on the other hand, should not be as strongly affected since mostly triples or higher order
systems eject companions, always leaving a lower order multiple. To disrupt a binary,
a stellar encounter is generally needed to provide the third body.

\begin{figure}
\plotone{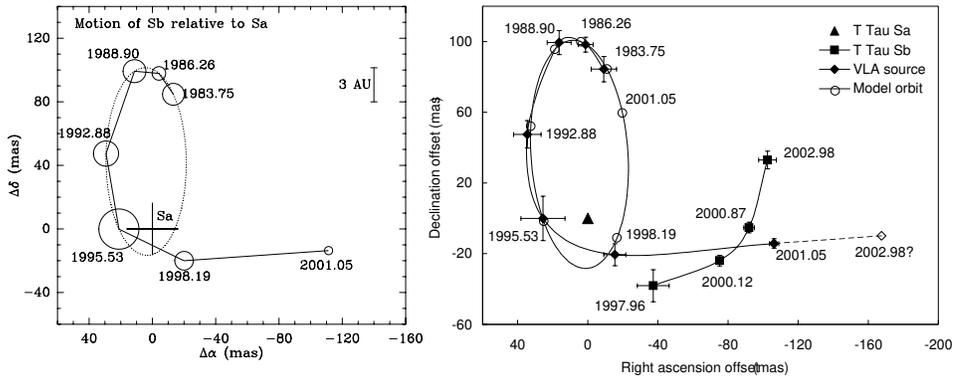}
\caption{Orbital evolution of the T~Tau~S system. The left panel shows the centimeter
emission feature relative to T~Tau~Sa, as found by VLA (from Loinard, Rodr\'{\i}guez,
\& Rodr\'{\i}guez 2003). After 1997, there is an apparent change of orbital path,
interpreted as a possible ejection of T~Tau~Sb. The right panel is from Furlan et al.\ 
(2003) and adds recent AO observations that show this centimeter feature not to be
associated with T~Tau~Sb, but possibly emanating from a fourth component being
ejected.
\label{fig-3}}
\end{figure}

An interesting case where we may be observing a current ejection is the prototype system
of all T~Tauri stars, T~Tauri. Loinard, Rodr\'{\i}guez, \& Rodr\'{\i}guez (2003) used
archival Very Large Array (VLA) data
to follow the orbital evolution of a centimeter radio emission feature
coincident with one of the southern
components in the triple system, T~Tau~Sb. They derive an orbit of Sb around the unseen 
(in their data) component Sa, that from 1983 to 1997 was well described by an ellipse, but
since then has departed significantly from the initial orbit (Fig.~3a). Furlan et al.\ 
(2003) dispute the identification of the radio source with the 
T~Tau~Sb companion, using recent AO NIR imaging. By following the evolution of the NIR
component since 1997 (when it was first noted in speckle data by Koresko 2000), Furlan
et al.\ (2003) derive a significantly different orbit from that of the radio emission
(Fig.~3b). They argue that the centimeter emission probably comes from the ejection of a
{\it fourth} component in the T~Tau system, still unseen in the AO NIR data. Statistically,
to observe an ejection in action is very unlikely, so we are extremely fortunate if
this is indeed an ejection. Future AO observations in the coming decade will clarify this.

\section{Summary and outlook}
Adaptive optics systems of today provide, for the first time, high enough angular
resolution to determine dynamical masses of the nearest T~Tauri multiple systems.
A few preliminary estimates of masses have already been reported in the literature.

The coming years will see an increase both in the number of systems with dynamical
mass determinations, and in the accuracy of the estimates. Looking beyond the nearest
systems, optical interferometry, with notably VLTI and the Keck interferometer,
will resolve binaries ten times more distant than AO alone.

By resolving multiple systems, stars can be accurately put on col\-our-mag\-ni\-tude
diagrams and compared to theoretical early evolutionary tracks. In combination with
accurate dynamical mass estimates, models will be much better constrained and
calibrated for use in, e.g., estimating the initial mass function for different
star forming regions.

\begin{figure}
\plotone{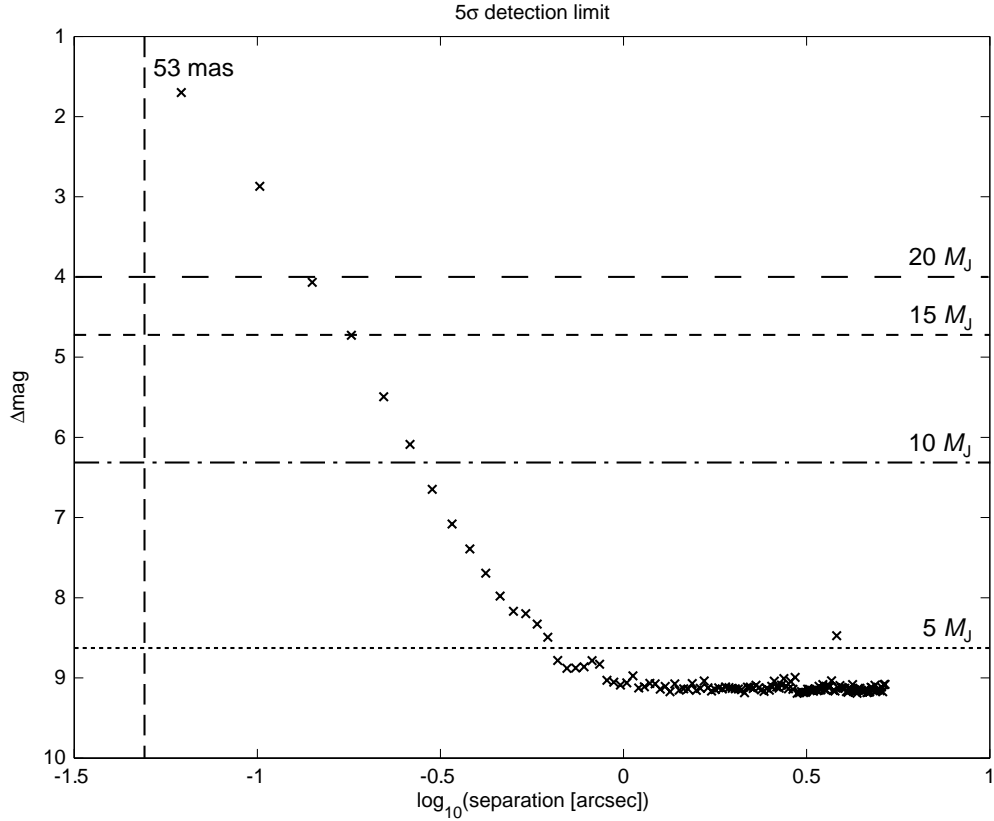}
\caption{VLT/NACO $5\sigma$ contrast sensitivity limit in $H$-band as a function of separation
for the binary $\eta$~Cha~9 of the $\eta$~Chamaeleontis association (Mamajek, Lawson, \&
Feigelson 1999). The vertical dashed line shows the diffraction limit, and the horizontal
lines show expected flux ratios of sub-stellar companions at the distance ($\sim$100~pc)
and magnitude ($H=10$) of $\eta$~Cha~9. Luminosities of the sub-stellar companions are
derived from Chabrier et al.\ (2000) and Baraffe et al.\ (2003), assuming an age of 10~Myr.
The exposure time was $2190\times0.34$~s. At small separations the noise is dominated by
speckle noise from the point spread function, while the sky-noise and read-out noise
dominate at larger separations. From Jayawardhana et al.\ (2003).
\label{fig-4}}
\end{figure}

High angular resolution observations in combination with spectroscopic searches
have the potential to carry out a complete census of companions in nearby young
clusters. Fig.~4 shows an example of the sensitivity to companions for a
state-of-the-art AO system. Note that AO systems now are sensitive enough
to probe the ``brown dwarf desert'' around young stars, as young brown dwarfs are
much more luminous than older ones (Baraffe et al.\ 2003). By increasing the statistics
and improving the observational bias corrections, the multiplicity and companion 
fraction as a function of separation and age may be firmly established, giving
important clues to the formation of multiples and stars in general.

\acknowledgements
I would like to thank the organisers for inviting me to give this short review,
C.F. Liljevalch j:or for a travel grant to get to Sydney, my supervisor Ren\'{e} Liseau
for supporting my participation despite being close to the deadline of my Ph.D.
defence, and Ray Jayawardhana and G\"{o}sta Gahm for valuable comments and suggestions.

\end{document}